\documentclass[manuscript,screen]{acmart}

\AtBeginDocument{%
 \providecommand\BibTeX{{%
 \normalfont B\kern-0.5em{\scshape i\kern-0.25em b}\kern-0.8em\TeX}}}

\copyrightyear{2025}
\acmConference[arXiv]{arXiv}{2025}{}

\usepackage{amsmath}
\usepackage{booktabs}
\usepackage{multirow}
\usepackage{tabularx}
\usepackage{subcaption}
\usepackage{graphicx}
\usepackage{url}

\begin{document}

\title{A Tool for Estimating Success Rates of Raycasting-Based Object Selection in Virtual Reality}

\author{Tatsuya Okuno}
\affiliation{\institution{Meiji University}\city{Tokyo}\country{Japan}}

\author{Haruto Shimizu}
\affiliation{\institution{Meiji University}\city{Tokyo}\country{Japan}}

\author{Nobuhito Kasahara}
\affiliation{\institution{Meiji University}\city{Tokyo}\country{Japan}}

\author{Taiyu Honma}
\affiliation{\institution{Meiji University}\city{Tokyo}\country{Japan}}

\author{Shota Yamanaka}
\email{syamanak@lycorp.co.jp}
\orcid{0000-0001-9807-120X}
\affiliation{\institution{LY Corporation}\city{Tokyo}\country{Japan}}

\author{Homei Miyashita}
\affiliation{\institution{Meiji University}\city{Tokyo}\country{Japan}}

\renewcommand{\shortauthors}{Okuno et al.}
\renewcommand{\shorttitle}{Success-Rate Estimation Tool for VR Raycasting}

\begin{abstract}
As XR devices become widespread, 3D interaction has become commonplace, and UI developers are increasingly required to consider usability to deliver better user experiences.
The HCI community has long studied target-pointing performance, and research on 3D environments has progressed substantially.
However, for practitioners to directly leverage research findings in UI improvements, practical tools are needed.
To bridge this gap between research and development in VR systems, we propose a system that estimates object selection success rates within a development tool (Unity).
In this paper, we validate the underlying theory, describe the tool's functions, and report feedback from VR developers who tried the tool to assess its usefulness.
\end{abstract}




\maketitle

\section{Introduction}
Virtual reality (VR) technology provides high realism and interactivity that cannot be achieved with conventional 2D displays.
In VR environments, selecting user-interface (UI) elements with a controller is one of the most fundamental operations underlying almost all interactions.

Pointing in VR is known to be more complex than in 2D environments due to factors unique to 3D interaction, such as target layouts along the depth dimension~\cite{Amini20253DFittsReview}.
In particular, the difficulty of selecting targets that are far from the user or appear small increases substantially.
This can lead to operational errors and frustration, ultimately degrading the overall quality of the VR experience.
Therefore, if VR application developers and designers can quantitatively evaluate how easy it is to select UI elements, they may be able to improve VR experiences.

To achieve this goal, we build a mathematical model that estimates the success rate of pointing operations in VR and develop a UI analysis tool based on the model.
Our model complements Fitts' law~\cite{Fitts1954Information}, which is widely known for predicting movement time, by focusing on success rates.
For success-rate estimation, we leverage insights from EDModel~\cite{Yu2019}, an endpoint-distribution model specialized for VR.
Ultimately, by providing our approach in a practical form similar to the 2D UI analysis tool \textit{Tappy}~\cite{usuba24arxiv,Yamanaka24arXivFigmaTappy}, we aim to enable the application of theoretical findings in HCI to real-world VR application development.

We provide our target-pointing success rate estimation tool as an open-source project on GitHub (\url{https://github.com/Fvkuinu/RayCastSuccessPredictor}).
Readers can install and use it as a plugin within the Unity engine.

\section{Related Work}
\label{relatedWork}

\subsection{Pointing in VR Space}
Various selection techniques exist for pointing in VR, and raycasting (extending a ray from a controller to select an object) is widely adopted because it enables the manipulation of distant objects with low physical effort.
However, raycasting suffers from reduced accuracy when targets are small and/or far away~\cite{Argelaguet2013Survey}.
To address this issue, many interaction techniques that assist pointing have been proposed~\cite{Lu20bubble,Vanacken2007BubbleCursor,Bateman13bubble,Baloup19bubble}.
In parallel, modeling users' pointing behavior to understand and predict performance has also been considered important~\cite{Kopper2010MotorBehavior}.
Many studies have used Fitts' law~\cite{Fitts1954Information} to estimate movement time.

\subsection{Success Rate of Pointing}
Early studies by Crossman and Welford showed that endpoint distributions follow a Gaussian distribution in 1D pointing tasks~\cite{Crossman56,welford1968fundamentals}.
Subsequent work on 2D pointing assumed that endpoint distributions follow a bivariate Gaussian distribution and computed the success rate of pointing, transforming it into an index of difficulty ($\mathit{ID}$) of Fitts' law~\cite{Grossman2005Probablility}.
Other work constructed error-rate models for mouse pointing based on effective target width and demonstrated applicability to both 1D bar targets and 2D circular targets~\cite{Wobbrock2008ErrorModel, Wobbrock2011ErrorModel2D}.

Bi et al.\ proposed a Dual Gaussian Distribution Model based on the dual-distribution hypothesis~\cite{Bi2013FFitts}, which explains tap-point distributions as a mixture of two independent Gaussian distributions: one that varies with target size and another that reflects an absolute distribution due to factors such as finger thickness and motor precision~\cite{Bi2016}.
Models extending this idea to various target shapes and conditions (e.g., 3D, moving targets) have also been proposed~\cite{Usuba20221Dto2D, Huang2020Ternary-Gaussian, Huang20182D1DTernary-Gaussian, Huang20192DTernary-Gaussian, Zheng20253DTernary-Gaussian, yamanaka2020servo}.

Among these, EDModel~\cite{Yu2019} is specialized for endpoint distributions in VR pointing.
This model investigated how target width ($W$), movement amplitude ($A$), and VR-specific depth ($Z$) affect endpoint distributions.
It showed that endpoint distributions in VR can also be approximated very well by a bivariate Gaussian distribution, and it derived regression equations that predict the distribution parameters (mean and covariance) from task-condition terms such as $W$ and $A$.
The predicted endpoint distribution follows a bivariate Gaussian distribution, as shown in Equation~\ref{eq:dual_Gaussian}.
\begin{equation}
\mu =
\begin{bmatrix}
\mu_x \\
0
\end{bmatrix}
,\quad
\Sigma =
\begin{bmatrix}
\sigma_x^2 & 0 \\
0 & \sigma_y^2
\end{bmatrix}
\label{eq:dual_Gaussian}
\end{equation}
Given the predicted distribution, the selection success rate $\mathit{SR}$ for a target region $D$ can be computed as the probability mass within $D$:
\begin{equation}
\mathit{SR} = \iint_D \frac{1}{2\pi\sigma_x\sigma_y} \exp\left( - \left( \frac{(x - \mu_x)^2}{2\sigma_x^2} + \frac{y^2}{2\sigma_y^2} \right) \right) \,dx\,dy
\label{eq:final_integral}
\end{equation}
However, the prior work~\cite{Yu2019} did not examine how the estimation accuracy of success rate changes depending on combinations of terms included in the model.
Therefore, in this study, we predict endpoint distributions using models with different term sets and compare their success-rate estimation accuracy.

\subsection{UI Design Support Tools}
In 2D UI design, the development of practical support tools based on quantitative predictive models has been active.
\textit{Tappy}~\cite{usuba24arxiv,Yamanaka24arXivFigmaTappy} is a suite of tools for estimating tap success rates of UI elements in smartphone webpages and applications, theoretically grounded in the Dual Gaussian Distribution Model.
Tappy accurately identifies tappable regions that are difficult to judge from appearance alone (e.g., whether an area around text accepts tap events) and visualizes success rates.
Evaluations using Tappy suggest that it helps developers and designers discover UI elements with lower-than-expected success rates and implementation issues, leading to concrete design fixes and data-driven decision-making.
A browser extension version, \textit{Tap Analyzer}~\cite{LIFULL24tapAnalyzer}, has also been provided.

Our goal is to realize a success-rate estimation tool that extends this concept to VR.
Because VR differs from 2D UIs in that it introduces depth and commonly uses raycasting for selection, we revisit and analyze an endpoint-distribution-based success-rate model suitable for VR.
We also implement a tool that uses the model and investigate how it is used by VR developers.

\section{Data Collection Experiment for Building the Estimation Model}
Our goal is to realize a success-rate estimation tool for VR that extends the concept of \textit{Tappy}~\cite{usuba24arxiv,Yamanaka24arXivFigmaTappy}.
To this end, we conducted a user study to revisit EDModel~\cite{Yu2019} and to explore alternative models with new or simplified terms that may be useful for success-rate estimation in practice.

\subsection{Participants and Apparatus}
Eighteen university students aged 18--24 (mean: 21.0; 2 female, 16 male) participated in the experiment.
We used a Meta Quest 3 headset with standard Touch Plus controllers.
The experimental program was implemented in Unity.
We used a raycasting selection technique, and the trigger action was performed with a controller button.

\subsection{Task}
\begin{figure}[t]
    \centering
    \includegraphics[width=1\linewidth]{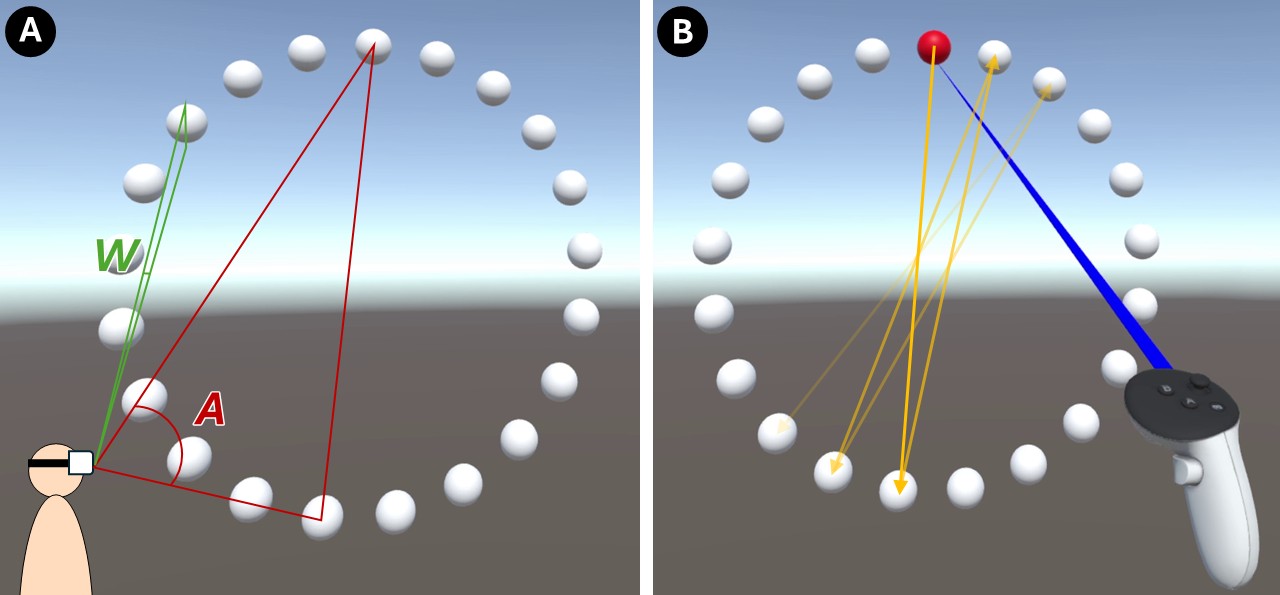}
    \caption{(A) Angular representations of target width $W$ and movement amplitude $A$.
    (B) Participants sequentially selected targets following the arrowed path.}
    \label{fig:studyParamExplaination}
\end{figure}

We conducted a pointing task in which participants sequentially selected 21 spherical targets arranged in a ring, similar to the multidirectional tapping task in ISO~9241-411~\cite{soukoreff2004towards} (Figure~\ref{fig:studyParamExplaination}A).
Participants selected targets in the order shown in Figure~\ref{fig:studyParamExplaination}B.
After a selection, a short sound was played, a new target was highlighted, and the next trial started.
Participants then moved the pointer to the next highlighted target.
Following prior work~\cite{Yu2019}, we did not provide feedback indicating whether the selection was correct.

We manipulated target width $W$ and movement amplitude $A$ (see Figure~\ref{fig:studyParamExplaination}A).
These variables were described in angular form~\cite{Kopper2010MotorBehavior, petford2018pointing}.
Specifically, we computed the angles corresponding to $W$ and $A$ from each participant's viewpoint.

We evaluated endpoint errors in an angle-based coordinate system rather than in physical distance.
In general, a point in 3D space can be represented using spherical coordinates with distance $r$ and two angles $(\theta, \phi)$.
Because a controller-emitted ray can be regarded as having infinite length, we can ignore the $r$ component and simplify endpoint error representation into a 2D coordinate using only the two angular components.

Accordingly, we define an endpoint $p$ as a 2D angular coordinate $p=(x,y)$.
Here, $x$ represents the angular error parallel to the movement direction (the straight line from the start target to the goal target), and $y$ represents the angular error perpendicular to that direction.
The origin is set at the center of the goal target, and both $x$ and $y$ can take negative values.

\subsection{Design and Procedure}
We employed an $8 \times 3$ within-subject design with two factors: target width $W$ and movement amplitude $A$.
Target width $W$ had eight levels, ranging from $1.0^\circ$ to $4.5^\circ$ in $0.5^\circ$ steps, and movement amplitude $A$ had three levels ($30^\circ$, $35^\circ$, and $40^\circ$).
These values were determined based on prior work~\cite{Yu2019} and pilot studies, and we confirmed that targets on the ring did not overlap even under the most extreme condition ($W=4.5^\circ$ and $A=30^\circ$).
The distance between participants and each target was fixed to a constant value (100\,m).
The presentation order of $W$--$A$ combinations was randomized per participant.
Following prior work~\cite{Yu2019}, we discarded the first trial for each $W$--$A$ combination (i.e., each ring), leaving 20 trials per condition for analysis.
In total, we collected 8,640 endpoints ($8\,W \times 3\,A \times 18\,\mathrm{participants} \times 20\,\mathrm{trials}$).

Participants first received an explanation of the VR device and the selection task.
They then wore the VR headset, performed a calibration for the forward direction, and completed practice trials before proceeding to the main trials.
After the experiment, participants completed a questionnaire to collect age, gender, and handedness.

We instructed participants to select targets at their comfortable, natural pace to record more natural selection behavior.
We also informed them that they could freely take short breaks if needed.

\section{Results}
\subsection{Data Screening and Normality Tests}
We obtained data from 8,640 trials.
Following prior work~\cite{Yu2019}, we removed outliers for each participant and each target condition: trials where the first click coordinate were more than 3 standard deviations away from the center on either axis.
We also applied the same 3$\mathit{SD}$ rule for the movement time.
As a result, 313 trials were excluded.
We grouped the remaining data by participant and condition for analysis.

For all endpoint coordinates, the Kolmogorov--Smirnov test confirmed normality on both the $x$ and $y$ axes ($\alpha = 0.05$).
Next, for each participant and target condition, we used maximum likelihood estimation to estimate the Gaussian mean $\mu$ and standard deviation $\sigma$ for both axes.
We also computed the correlation $\rho$ between the two axes.
Using the five dependent variables $\mu_x$, $\sigma_x$, $\mu_y$, $\sigma_y$, and $\rho$, we computed the mean vector $\boldsymbol{\mu}$ and covariance matrix $\boldsymbol{\Sigma}$ (see Equation~\ref{eq:dual_Gaussian}).

\subsection{Endpoint Distributions}
\label{endpointDistribution}
We performed repeated-measures ANOVAs (RM-ANOVAs) for the five dependent variables.
Degrees of freedom were adjusted using the Greenhouse--Geisser correction.
$W$ had significant main effects on $\mu_x$ ($F_{2.81}=9.95$, $p<0.001$, $\eta_p^2=0.369$), $\sigma_x$ ($F_{6.91}=56.30$, $p<0.001$, $\eta_p^2=0.768$), and $\sigma_y$ ($F_{2.94}=40.90$, $p<0.001$, $\eta_p^2=0.706$).
No other main effects or interactions were observed.

For dependent variables with significant main effects, we conducted linear regression analyses.
The results showed linear relationships between $W$ and $\mu_x$ ($R^2=0.73$), between $W$ and $\sigma_x$ ($R^2=0.98$), and between $W$ and $\sigma_y$ ($R^2=0.89$).

Based on these results, we constructed the following bivariate Gaussian distribution:
\begin{equation}
\mu =
\begin{bmatrix}
 eW + f \\
 0
\end{bmatrix},
\quad
\Sigma =
\begin{bmatrix}
 (aW + b)^2 & 0 \\
 0 & (cW + d)^2
\end{bmatrix}
\label{eq:dual_GaussianWithParam}
\end{equation}
where the constants from regression are $a=0.1102$, $b=0.23130$, $c=0.0715$, $d=0.2311$, $e=-0.0623$, and $f=-0.0846$.

\subsection{Success-Rate Estimation}
\begin{figure}[t]
    \centering
    \includegraphics[width=0.64\linewidth]{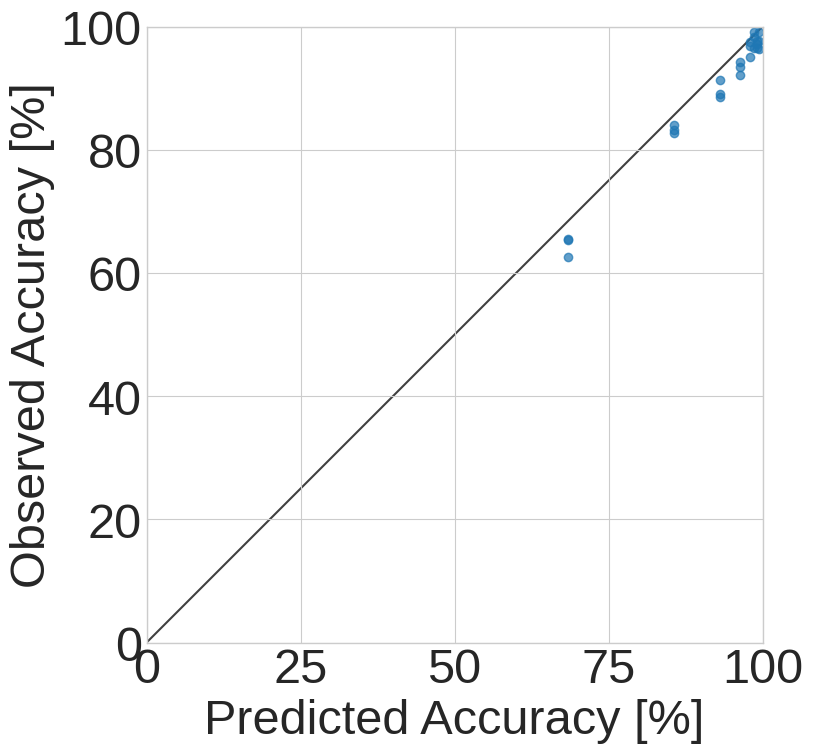}
    \caption{Observed success rates for spherical targets and success rates estimated using the endpoint distribution derived from our experiment.}
    \label{fig:SRplot}
\end{figure}

Next, using the bivariate Gaussian distribution derived from our experiment (Equation~\ref{eq:dual_GaussianWithParam}), we estimated target selection success rates for each condition based on Equation~\ref{eq:final_integral}.
Figure~\ref{fig:SRplot} shows the estimated and observed success rates.
The maximum difference between observed and estimated success rates was 5.82\% ($W=1.0^\circ$, $A=30^\circ$).
The mean absolute error ($\mathit{MAE}$) was 2.35\%, and $R^2$ was 0.987.
With leave-one-out cross-validation (LOOCV), we obtained $\mathit{MAE}=2.39\%$ and $R^2=0.985$.

\section{Additional Analyses for Model Refinements}
To explore the applicability and practicality of our model for tool usage, we conducted additional analyses focusing on three key factors.
Our goal here is to clarify an appropriate model configuration from practical perspectives such as computational cost and usability when applied to a success-rate estimation tool.
Specifically, we examine the validity of the model from the following three perspectives.

\subsection{Analyzed Factors}

\textbf{Using movement amplitude ($A$).}
First, we examine whether it is necessary to use the movement amplitude $A$ in the model.
Prior work using hand-based raycasting~\cite{Yu2019} included $A$ in the endpoint distribution prediction model.
However, in our task, the RM-ANOVA suggested that its effect was limited.
If a model without $A$ retains sufficient accuracy, success-rate estimation can be performed by considering only target size, which would improve usability because developers can use the tool without accounting for the pointer--target distance.
We thus compare models with and without $A$ to determine whether we can simplify the model without sacrificing estimation accuracy.

\textbf{Simplifying the offset ($\mu_x$).}
Second, we evaluate whether assuming $\mu_x=0$ is valid.
Many basic probabilistic models based on Fitts' law for pointing errors assume that users always aim at the target center, setting the mean error ($\mu_x,\mu_y$) to zero~\cite{Bi2016,Yamanaka20issFFF,Wobbrock2008ErrorModel}.
We compare a model without $\mu_x$ and a model where the mean error dynamically changes with $W$.
Through this comparison, we examine whether the simpler assumption is practically sufficient.

\textbf{Adopting a world-coordinate system.}
Third, we examine whether replacing the movement-direction-based coordinate system (movement direction as $x$, and the perpendicular direction as $y$) with Unity's standard world-coordinate system remains effective.
The movement-direction-based coordinate system has been used in prior work~\cite{Yu2019} and is suitable for motion analysis.
In contrast, a world-coordinate system may allow modeling effects such as pointer drift caused by forces applied at button press, including real-world influences like gravity and inertia.
Moreover, if success-rate estimation is possible in a world-coordinate system, developers would not need to consider varying elements such as pointer movement direction or target orientation.
Then, success rates could be computed based only on target size, simplifying implementation and reducing computational cost.

\subsection{Results}
We performed the three additional analyses and followed the same analysis procedure as in the main experiment.
Table~\ref{tab:ablation_loocv} shows success-rate estimation results for each model.
Compared with the baseline model, estimation accuracy decreased slightly, but $\mathit{MAE}$ remained around 2--4\% for all models.
In addition, the difference in $\mathit{AIC}$ from the baseline was less than 10, indicating that all models can provide practically useful estimation.
Therefore, depending on application contexts, each model can be used to improve utility, simplify the model, and reduce implementation complexity and computational cost while keeping accuracy within an acceptable range.

\begin{table*}[t]
    \centering
    \caption{Comparison of success-rate estimation accuracy across models in the additional analyses and generalization performance evaluated by LOOCV.}
    \label{tab:ablation_loocv}
    \begin{tabular*}{0.9\textwidth}{@{\extracolsep{\fill}}lccccr}
        \toprule
        & \multicolumn{2}{c}{Training data} & \multicolumn{2}{c}{LOOCV} & \\
        \cmidrule(lr){2-3} \cmidrule(lr){4-5}
        Model & $\mathit{MAE}$ (\%) & $R^2$ & $\mathit{MAE}$ (\%) & $R^2$ & $\mathit{AIC}$ \\
        \midrule
        Baseline & 2.354 & 0.987 & 2.394 & 0.985 & 82.50 \\
        With distance parameter ($A$) & 2.352 & 0.985 & 2.405 & 0.982 & 85.57 \\
        Offset simplification ($\mu_x=0$) & 3.907 & 0.982 & 3.933 & 0.982 & 90.01 \\
        World-coordinate system & 3.827 & 0.989 & 3.871 & 0.988 & 78.91 \\
        \bottomrule
    \end{tabular*}
\end{table*}

\begin{figure}[t]
    \centering
    \includegraphics[width=1\linewidth]{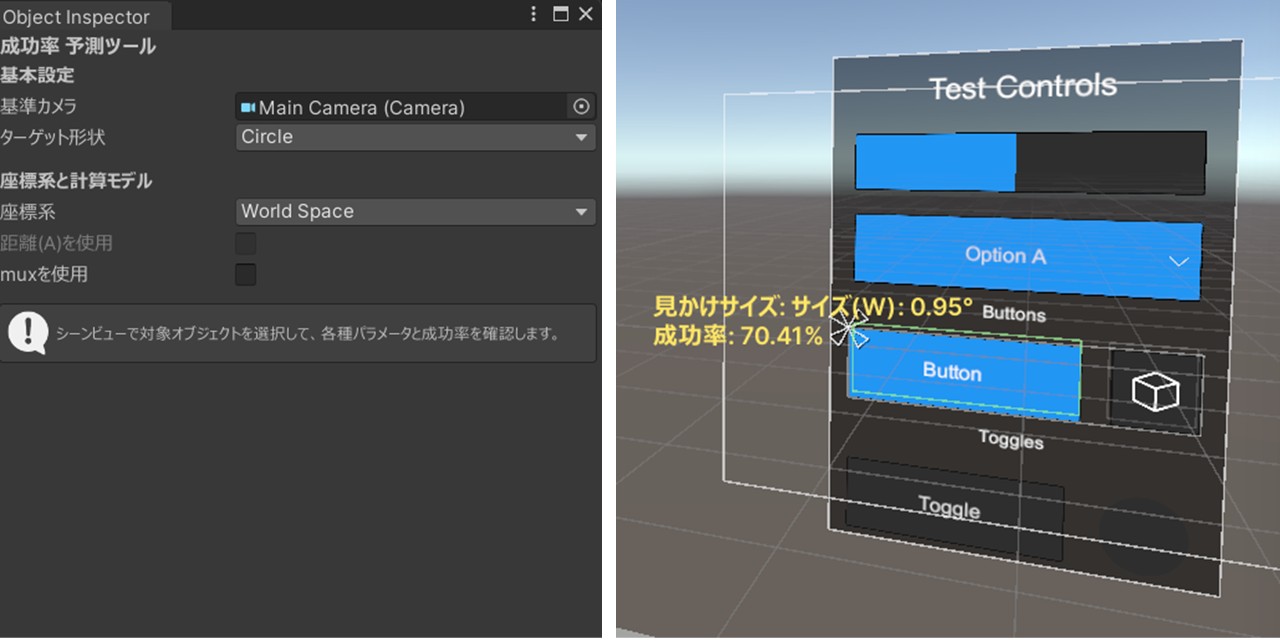}
    \caption{Left: Inspector window of the proposed success-rate estimation tool implemented as a Unity editor extension. Developers specify analysis settings such as the reference camera, target shape (e.g., circular), coordinate system (e.g., world space), and optional model parameters. Right: Unity scene view showing a VR UI canvas. When a UI element is selected, the tool overlays its apparent angular size (target width $W$ in degrees, e.g., $0.95^\circ$) and the predicted selection success rate (e.g., 70.41\%) near the object.}
    \label{fig:VRTappy}
\end{figure}

\section{Success-Rate Estimation Tool}
To enable VR application developers to use the success-rate estimation model easily, we implemented a UI analysis tool.
This tool applies the concept of the smartphone UI analysis tool \textit{Tappy}~\cite{usuba24arxiv,Yamanaka24arXivFigmaTappy} to VR systems.

The tool is implemented as a Unity editor extension.
During VR scene design, designers and developers can select an arbitrary UI element (e.g., buttons or panels) and confirm its predicted selection success rate in real time (Figure~\ref{fig:VRTappy}).
The tool automatically obtains the size of the selected UI, its distance from the camera (depth), and an assumed pointing movement amplitude, and then computes and displays the success rate based on the estimation model.
This enables developers to quantitatively identify usability issues early, such as ``this button is too far to click easily'' or ``this menu is too small and may cause selection errors,'' and to improve them.

\subsection{Implementation}
Based on prior work~\cite{usuba24arxiv,Yamanaka24arXivFigmaTappy} and characteristics of VR environments, we describe key features of our tool for estimating selection success rates of VR UI elements.

\textbf{Target-shape selection.}
Developers can choose the target shape as either a sphere or a rectangle.
Our experiment used spherical targets, so spheres are the default supported shape in the model.
We did not directly conduct experiments on rectangular targets in this study.
However, because we observed no correlation between the $x$- and $y$-axis endpoint distributions, we implemented success rate estimation for rectangular targets by assuming independence across axes.

In addition, for rectangles, the effective width and height can vary depending on pointer movement direction.
To address this, we used the insight from the additional analysis that estimation remains possible in a world-coordinate system and introduced a method that computes rectangle dimensions in an absolute coordinate system.

\textbf{Accounting for selection-center offset.}
Developers can choose whether to assume that the pointing ``selection center'' is the geometric center of the target.
In our experiment, we observed a tendency for the selection center to shift toward the user (opposite the movement direction) as target size increases.
On the other hand, the additional analysis showed that the model remains effective even when assuming the target center as the selection center.
Therefore, the tool provides an option to enable/disable this offset assumption depending on developers' needs.

\textbf{Accounting for distance to the target.}
We provide an option to include or exclude the pointer-to-target distance in the success-rate computation.
The additional analysis showed that including distance does not produce a large difference in estimation accuracy.
However, our results also suggest a slight tendency for success rates to decrease as distance increases.
Thus, we provide this option for developers who want to account for that effect.

\subsection{How to Use}
First, a developer selects a reference camera and a target object whose success rate they want to estimate.
Then, the tool displays the target's size and predicted selection success rate around the target in the editor (Figure~\ref{fig:VRTappy}, right).
In addition, developers can select the options described above in the tool window (Figure~\ref{fig:VRTappy}, left) to choose the model configuration best suited to their purpose.

\begin{figure}[t]
    \centering
    \includegraphics[width=1\linewidth]{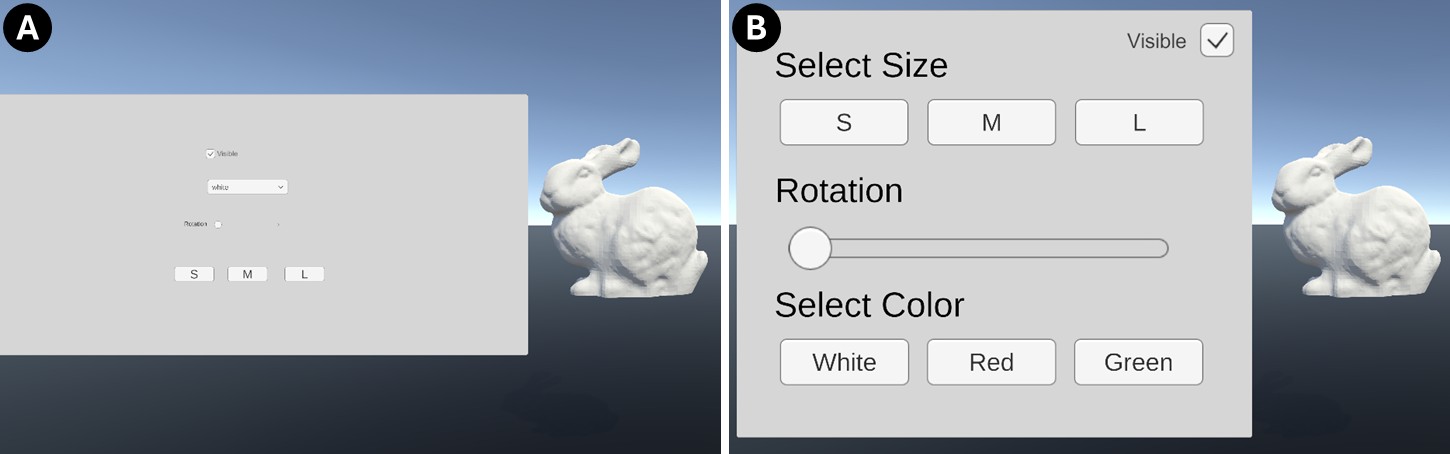}
    \caption{(A) UI before adjustment. (B) UI after adjustment by a participant. The canvas was attached to the left-hand controller.}
    \label{fig:taskBeforeAfter}
\end{figure}

\section{User Study with VR Developers}
To examine the usefulness of our success-rate estimation tool, we conducted a user study with two participants who had VR game development experience.
Considering their backgrounds, we asked each participant to try the tool and provide feedback in a different format.
Because this was a small-scale study with only two experienced developers, it does not comprehensively evaluate the proposed system.
Instead, our goal was to obtain qualitative insights for future improvements.

\subsection{Controlled-Setting User Study}
One participant had about one year of VR application development experience but was not currently engaged in it.
Therefore, rather than evaluating the tool in a real development environment, we investigated its usefulness in a controlled task.

We first explained the basic operations of the tool (selecting a UI element and checking its success rate).
As a task, we presented a sample UI for adjusting the color and size of a 3D model (a rabbit) (Figure~\ref{fig:taskBeforeAfter}A) and asked the participant to ``improve the UI so that it is easier to operate, referring to the selection success rates displayed by the tool.''
Figure~\ref{fig:taskBeforeAfter}B shows the UI after the participant completed the task.
Afterward, we interviewed the participant about how the tool helped in the UI improvement process and what practical issues exist.

During the task, the participant first checked each button and slider on the UI panel and judged their functional importance.
For example, the participant considered buttons for changing color to be frequently used, while a button for toggling the 3D model visibility was considered less frequently used.
Then, the participant used the tool to confirm success rates numerically for each UI element.
When the participant found a frequently used button with a low predicted success rate, they enlarged the element while monitoring the displayed success rate.
For a set of high-importance buttons, we observed that the participant set a concrete goal such as ``increase button size until the tool shows a success rate of 95\% or higher'' and proceeded accordingly.

The participant stated that it was useful to analyze success rates and modify UI elements without repeatedly wearing the HMD and testing in VR.
They also noted that the tool provides an objective numeric indicator that helps identify issues that might otherwise be overlooked due to developers' subjective judgments or habituation.
As an expected usage context, the participant suggested that being able to confirm numbers such as success rate before the quality assurance (QA) phase could serve as a design guideline when planners and engineers discuss UI operability.

\subsection{Real Development Environment User Study}
The other participant had five years of VR system development experience and was a VRChat world creator.
We asked the participant to use the tool for one week and provide feedback via a questionnaire.
The questionnaire asked about usability, feasibility of adopting the tool in future workflows, and challenges in their usual UI development practice, mainly through open-ended responses.

The participant reported that they typically adjusted UI size and placement based on subjective criteria such as experience and intuition from past projects.
They found it useful that our tool provides an objective indicator for UI design decisions that had previously relied heavily on intuition.
In particular, they considered the tool useful in early stages when deciding UI sizes and stated that they would like to integrate it into their development workflow.
They also expected that the tool could reduce variations in design decisions across developers and help reduce UIs that look acceptable but are actually hard to operate.

This study was conducted in the context of VRChat UI development, where technical constraints exist: due to the VRChatSDK specification, debugging on real devices is difficult within the Unity editor.
The participant indicated that our tool could mitigate this issue by enabling usability evaluation without going through VR device debugging.
Furthermore, in VRChat, users' avatar heights vary widely, resulting in different viewpoints across users.
Designing UIs that are accessible for users of many heights is challenging.
The participant suggested that our tool could help quantify how height differences affect UI accessibility without repeatedly debugging with multiple height models, thereby potentially contributing to addressing this challenge.

\section{Discussion}
In this study, we constructed models under several assumptions, and further work is needed to ensure generalizability.
Such generalizability is essential for applying the tool to diverse UIs.

First, there are limitations in the experimental setting that forms the basis of our model.
Our experiment used spherical targets placed on a ring at an equal distance from the user.
In practical UIs, however, targets are often placed on planes whose distance and visual angle vary dynamically.
In addition, pointing accuracy may differ between central and peripheral visual fields.
Therefore, careful consideration is required regarding prediction accuracy when applying our model to such conditions.
Moreover, in the tool, we assumed that endpoint distributions for each axis can be estimated independently and thus support rectangles.
Validating this assumption and the model applicability under conditions different from our experiment remains an important future task.

Second, participant demographics were limited.
Participants were university students with a mean age of 21.0.
We did not consider changes in motor ability due to aging or the effect of VR experience on pointing accuracy.
Additional validation involving more diverse user populations (different age groups and VR proficiency levels) is needed to improve generalizability.

Third, the input device and selection method were limited.
Our study assumes a controller-based raycasting.
The physical shape of input devices can affect selection accuracy, as seen in phenomena such as the Heisenberg effect (pointer drift when pressing a button)~\cite{Wolf2020HeisenbergEffect}.
Thus, it is not guaranteed that our model will achieve similar accuracy for other devices or selection methods.
Applying our findings to other pointing techniques, such as virtual hands or eye tracking, requires separate investigation.

We also conducted a user study to examine the usefulness of the developed tool.
The results suggested that the tool can provide an objective indicator in UI design processes and trigger improvements.
For example, participants gave positive feedback such as ``I now have an objective guideline for UI sizes that I previously set by intuition.''
These findings suggest that our tool can help bridge the gap between HCI knowledge and development practice, especially in XR development (AR/VR) where real-device testing requires effort and individual differences such as height and posture affect UI operability.
However, our user study aimed to collect feedback for tool improvement and was limited by a small number of experienced participants.
A larger-scale, quantitative evaluation is required for comprehensive assessment.

As future work, we can extend the tool beyond editor usage.
For example, the tool could be used while wearing the HMD to display success rates from the current pointer position to each UI element.
We also plan to quantitatively evaluate usefulness by comparing UIs designed with and without the tool in terms of design time reduction and quality improvement.
Furthermore, beyond UI design support, our approach could be applied to action games as a tool for quantitatively designing and analyzing difficulty by adjusting the sizes and placements of objects such as blocks.

\bibliographystyle{ACM-Reference-Format}
\bibliography{sample-base}

\end{document}